\documentclass[twocolumn,showpacs,preprintnumbers,amsmath,amssymb,prb]{revtex4}
\usepackage{graphicx}
\usepackage{dcolumn}
\usepackage{bm}

\begin{document}
\title{QMC Calculation of the Electronic Binding Energy in a $C_{60}$ Molecule}
\author{Fei Lin, Jurij \v{S}makov, Erik S. S$\o$rensen, Catherine Kallin and A. John Berlinsky}
 \affiliation{Department of Physics and Astronomy, McMaster University, Hamilton, Ontario,
 Canada L8S 4M1}
 \date{\today}

\begin{abstract}
Electronic energies are calculated for a Hubbard model on the
$C_{60}$ molecule using projector quantum Monte Carlo (QMC).
Calculations are performed to accuracy high enough to determine
the pair binding energy for two electrons added to neutral
$C_{60}$.  The method itself is checked against a variety of other
quantum Monte Carlo methods as well as exact diagonalization for
smaller molecules.  The conclusion is that the ground state with
two  extra electrons on one $C_{60}$ molecule is a triplet and,
over the range of parameters where QMC is reliable, has a slightly
higher energy than the state with electrons on two separate
molecules, so that the pair is unbound.
\end{abstract}


\maketitle

\section{Introduction}
The discovery of superconductivity in the alkali-metal-doped bulk
fullerenes $K_3C_{60}$ and $Rb_3C_{60}$ \cite{hebard91, ross91}
sparked intense interest in fullerene-based materials, leading to
extensive experimental and theoretical studies (for a review see
Ref. \onlinecite{gunnarsson97}). Theoretical calculations to
explain the insulating and superconducting properties of bulk
fullerene materials fall into two major categories: molecular
level calculations which determine the effective interactions of
intramolecular electrons,\cite{kivelson91a, kivelson91b, joyes92,
lu94, sheng94, dong95, dong96, krivnov94} and lattice level
calculations based on an effective Hamiltonian in which the
intramolecular degrees of freedom have been integrated
out.\cite{granath0203, gunnarsson96, capone02}

Although much of the work has focused on the phonon mechanism for
superconductivity in the alkali-$C_{60}$'s, some authors have proposed a purely
electronic mechanism.  In particular,
it was argued by Chakravarty, Kivelson and co-workers (CK)
that electronic interactions within a single $C_{60}$ molecule
can lead to an effective attraction between charge carriers.
\cite{kivelson91a, kivelson91b, kivelson92a, kivelson92b,
kivelson01} This argument was supported by perturbative calculations
of the electronic binding energies of the conventional one band
Hubbard model, on the $C_{60}$ structure. Results of the CK
calculation suggest that electrons have a tendency to form paired
states in a single fullerene molecule, rather than remain
separate. This tendency could be the origin of the attractive
interaction which is an essential ingredient of the BCS theory of
superconductivity.

However, one might doubt the applicability of
perturbation theory to this problem. First, the Hubbard repulsion $U$ in the CK
calculation is approximately 75\% of the bandwidth, so it is
hardly a small parameter. Also, the binding energy is
typically a small quantity, calculated from the difference of large internal
energies of the $C_{60}$ molecule at different electron dopings.
Low order perturbation theory estimates of such subtle energy
differences may be unreliable. Thus it is interesting to repeat the
calculation using different methods, which might lend support to or cast doubt
upon the perturbation theory results.

\begin{figure}
  \begin{tabular}{c}
  \resizebox{80mm}{!}{\includegraphics{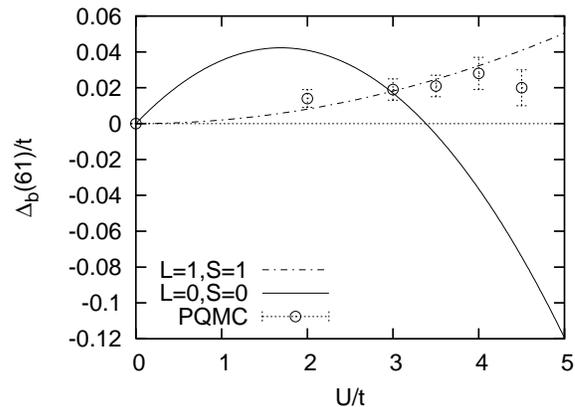}} \\
  \end{tabular}
  \caption{Comparison of electronic pair binding energies
$\Delta_{\textrm{b}}(61)/t$, defined in Eq. (\ref{pairbind}),
obtained from perturbation theory in different spin sectors (solid and
dash-dot lines) \cite{kivelson91a, kivelson91b, ostlund} and PQMC
calculations on a $C_{60}$ molecule.  PQMC finds $S=1$ for ground states with
62 electrons.}
  \label{perturb61}
\end{figure}
In this paper we use quantum Monte Carlo (QMC) calculations to
estimate the binding energy of pairs of electrons on a single $C_{60}$
Hubbard molecule. In order to establish a high level of confidence in
our results, we use a number of complementary QMC methods, including
auxiliary field QMC for both real and imaginary chemical potentials
\cite{dagotto90, note1} and stochastic series expansion (SSE) at
finite temperature $T$, and projector QMC (PQMC) at $T=0$ on a series of Hubbard
molecules, with the number of sites ranging from 4 to 60. In addition,
comparison is made to the results of exact diagonalization (ED) for
systems of up to 12 sites. Our main result, shown in
Figure \ref{perturb61}, is a comparison of the pair binding energy (see
below for a definition) for two electrons added to a neutral $C_{60}$
molecule, to the perturbation calculations of CK. All energies are
measured in units of the hopping parameter $t$. In contrast to
perturbation theory, which finds that the ground state is a singlet
for $U/t>3$, we find, in agreement with Hund's rule, that the ground
state remains a triplet state over the entire range of $U$
studied.  In particular, there is no indication of the attractive
singlet ground state which perturbation theory finds for $U/t$ greater
than about $3.3$.  Our QMC studies find small positive binding energies
for $U/t\le 4.5$, indicating that two separate molecules, each with
one extra electron, have lower energy than one molecule with two extra
electrons. The largest value of $U$ we are able to study is $4.5t$,
since the sign problem discussed below becomes unmanageable for larger
values of $U$.

The paper is organized as follows: in the next section we
introduce the model and the QMC methods used in our simulations.  We
then proceed to present the tests of QMC codes on smaller molecules
(truncated tetrahedron, etc.), where exact analytical or exact
diagonalization results are available. Then the results for the
$C_{60}$ molecule are presented and analyzed. Finally, a conclusion
based on our numerical results is drawn and the reliability of the
method is discussed.

\begin{table}
  \centering
  \begin{ruledtabular}
  \begin{tabular}{|l|lddd|}
      & $S_z$ &  \textrm{ED} &  \textrm{PQMC} & \textrm{sign} \\
    \hline
    $E_{10}$   &   0      & -14.506219 & -14.397(3) & 0.47 \\
    $E_{10}$   &   1      & -14.506219 & -14.504(2) & 1.00 \\
    $E_{11}$   &   1/2    & -13.623187 & -13.620(3) & 0.81 \\
    $E_{11}$   &   3/2    & -12.876242 & -12.880(4) & 0.57 \\
    $E_{12}$   &   0      & -12.697340 & -12.698(2) & 1.00 \\
    $E_{12}$   &   1      & -11.874844 & -11.856(3) & 0.52 \\
    $E_{13}$   &   1/2    & -10.701320 & -10.698(3) & 0.58 \\
    $E_{13}$   &   3/2    & -9.982385  & -9.969(3) & 0.46 \\
    $E_{14}$   &   0      & -8.725294  & -8.681(4) & 0.30 \\
    $E_{14}$   &   1      & -8.645244  & -8.643(4) & 0.54 \\
    \hline
    $\Delta_{1,0}$       & (1/2,0) &  0.996021 &  1.000(3) &  \\
    $\Delta_{1,0}$       & (3/2,0) &  1.714956 &  1.729(3) &  \\
    $\Delta_{-1,0}$      & (1/2,0) &  0.074154 &  0.078(3) &  \\
    $\Delta_{-1,0}$      & (3/2,0) &  0.821099 &  0.818(4) &  \\
    $\Delta_{\textrm{b}}(13)$  &  (0,0,1/2) & -0.019995 &  0.017(4) &  \\
    $\Delta_{\textrm{b}}(11)$  &  (0,1,1/2) &  0.042813 &  0.038(3) &  \\
  \end{tabular}
  \end{ruledtabular}
  \caption{Comparison of exact diagonalization and PQMC calculations
  on the truncated tetrahedron (12 sites) at $U=2t$. PQMC simulation
  parameters: $\beta=10/t$, $\Delta\tau=0.05/t$, $N_m=10^7$.
  $E_n(S_z)$ is the energy of a system with $n$ electrons and
  $z$-component of total spin $S_z$. $\Delta_{n,m}$ is the energy difference
  $E_{12+n}(S^n_z)-E_{12+m}(S^m_z)$ with $(S^n_z,S^m_z)$ given in the
  second column. For binding energies $\Delta_{\textrm{b}}(n)$ the second column
  shows $(S^{n+1}_z,S^{n-1}_z,S^{n}_z)$ -- the $S_z$ values for 3
  states involved in its calculation, in the order of appearance in
  Eq. (\ref{pairbind})}.\label{c12pqmc}
\end{table}

\section{Methodology}
Following CK, we consider a one-band Hubbard model with the
Hamiltonian $H=H_0+H_1$, defined on a $C_{60}$ molecule by
\begin{eqnarray}
H_0&=&-\sum_{\langle ij\rangle\sigma}t_{ij}(c_{i\sigma}^{\dagger}c_{j\sigma}+h.c.)-\mu\sum_{i\sigma}n_{i\sigma},\nonumber\\
H_1&=&U\sum_{i}n_{i\uparrow}n_{i\downarrow}-\frac{U}{2}\sum_{i\sigma}n_{i\sigma}.
\end{eqnarray}
Here $H_0$ contains the standard kinetic energy and chemical potential
term. The summation in the kinetic energy term is performed over all nearest neighbor pairs on a $C_{60}$ molecule. The hopping
constants $t_{ij}$ are chosen to be equal to $t$ for the single bonds, connecting a pentagon and a
hexagon, and to $t'=1.2t$ for the double bonds between two
hexagons. $H_1$ is a sum of the on-site Coulomb repulsion (Hubbard)
term and a diagonal term, added to make the model particle-hole
symmetric around $\mu=0$ on bipartite lattices. Clearly, this
additional term does not affect the value of the electronic binding
energy, which we choose to define as
\begin{equation}
\label{pairbind} \Delta_{\textrm{b}}(n)=E_{n+1}+E_{n-1}-2E_{n},
\end{equation}
where $E_n$ is the internal energy of a molecule with $n$
electrons. Note that this definition has the opposite sign,
compared to that of CK.\cite{kivelson91a} In our case the
tendency of the electrons to bind into pairs is indicated by a
negative value of the binding energy $\Delta_{\textrm{b}}$.

\begin{table}
  \centering
  \begin{ruledtabular}
  \begin{tabular}{|l|ccc|}
     &  &  $U=0$ & \\
     \hline
    $\mu$ & $-1$ & $0$ & $1$ \\
    \hline
    $n_{\textrm{exact}}$ &  0.7959095  &  1.0001765  &  1.2043146 \\
    $E_{\textrm{exact}}$ & -44.2672020 & -46.3708440 & -44.3478000 \\
    $n_{\textrm{AFQMC}}$ &  0.7959095  &  1.0001765  &  1.2043146 \\
    $E_{\textrm{AFQMC}}$ & -44.2672020 & -46.3708440 & -44.3478000 \\
    $S_{\textrm{AFQMC}}$ &  1.0       &  1.0       &  1.0       \\
    \hline
    \hline
     &  & $U=4t$ &  \\
     \hline
    $\mu$ & $-1$ & $0$ & $1$ \\
    \hline
    $n_{\textrm{SSE}}$   &  0.873(1)  &  1.0005(2)   &  1.126(1) \\
    $E_{\textrm{SSE}}$   & -16.5(2)   & -4.0(2)      &  13.9(1)  \\
    $S_{\textrm{SSE}}$   &  0.955     &  0.957       &  0.960    \\
    $n_{\textrm{AFQMC}}$ &  0.8734(2) &  1.000078(1) &  1.1266(2)\\
    $E_{\textrm{AFQMC}}$ & -16.61(2)  & -4.13(4)     &  13.74(6) \\
    $S_{\textrm{AFQMC}}$ &  1.0       &  1.0         &  1.0      \\
  \end{tabular}
  \end{ruledtabular}
  \caption{Comparison of the density $n$, total internal energy $E$
and average sign $S$ between exact analytical results $(U=0)$, SSE
$(U=4t)$ and AFQMC on a $C_{60}$ molecule. Simulation parameters:
$\beta=0.5/t$, $\Delta\tau=0.05/t$, $N_m=10^5$. In the SSE run $10^7$
measurements, separated by a full diagonal and directed loop 
update,\cite{sandvik9799,sandvik02} were performed. At lower 
temperatures SSE is unreliable due to the severe sign problem.} 
\label{c60afqmc}
\end{table}

\begin{figure*}
  \begin{tabular}{cc}
    \resizebox{75mm}{!}{\includegraphics{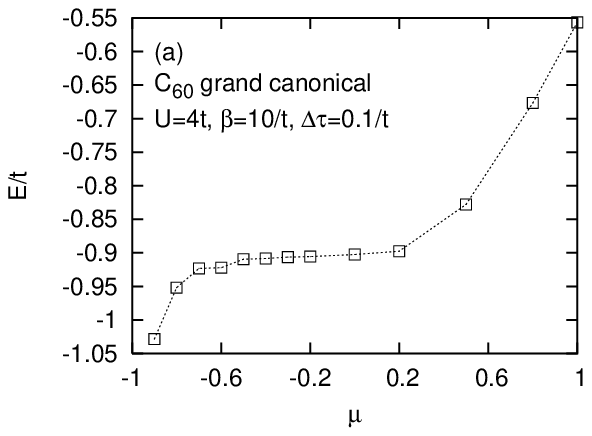}} &
    \resizebox{75mm}{!}{\includegraphics{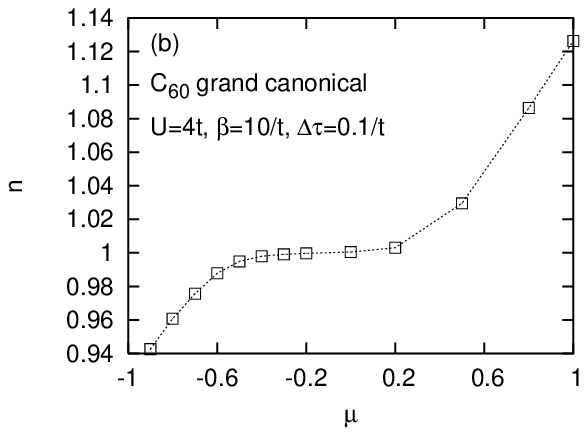}} \\
    \resizebox{75mm}{!}{\includegraphics{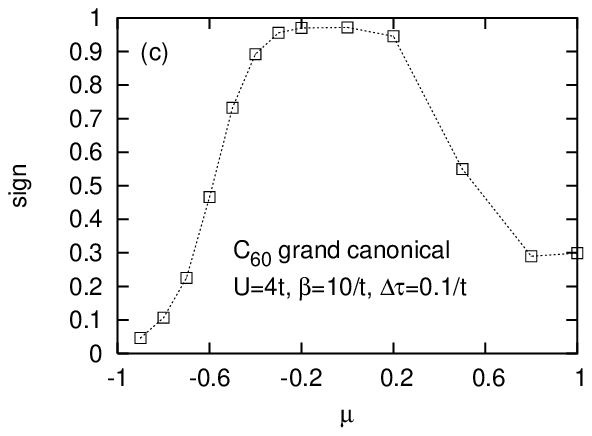}} &
    \resizebox{75mm}{!}{\includegraphics{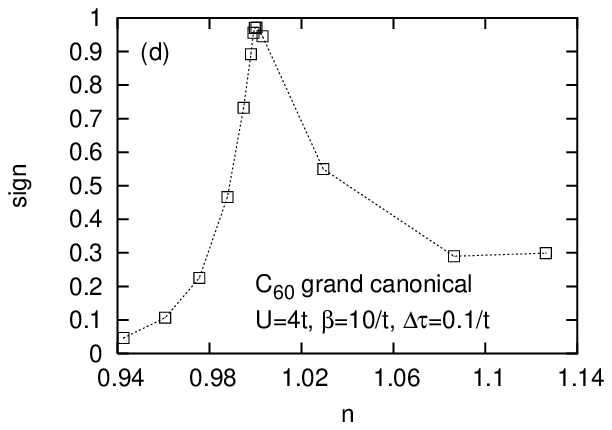}} \\
  \end{tabular}
  \caption{Grand canonical simulation of a $C_{60}$ molecule at
  various chemical potentials. Simulation parameters: $U=4t$,
  $\beta=10/t$, $\Delta\tau=0.1/t$, $N_t=10^3$, $N_m=10^5$.  We are
  only interested in the qualitative behavior of the system around
  half-filling, so the statistical errors are not estimated. (a)
  Energy per site vs. chemical potential. (b) Electron density
  vs. chemical potential.  (c) Average sign as a function of chemical
  potential. (d) Average sign as a function of electron
  density. Curves connecting the points are guides to the eye.}
  \label{c60gcs}
\end{figure*}

Determinant or auxiliary field QMC (AFQMC) has been widely used in
model Hamiltonian simulations since its introduction by
Blankenbecler \emph{et al.},\cite{sugar81a,sugar81b} and its
further development by Hirsch,\cite{hirsch83a83b, hirsch85} and
White \emph{et al.}\cite{white89} The application of this
technique to the one-band Hubbard model starts with the
Suzuki-Trotter discretization of the imaginary time in the grand
canonical partition function \cite{suzuki86}
\begin{eqnarray}
\label{zini}
Z_{\textrm{GC}}&=&\textrm{Tr}\,e^{-\beta H}\nonumber\\
&=&\textrm{Tr}\,e^{-\beta(H_{0}+H_{1})}\nonumber\\
&=&\textrm{Tr}\,\prod_{i=1}^{L}e^{-\Delta\tau(H_{0}+H_{1})},
\end{eqnarray}
where $\beta=1/(k_BT)$ is the inverse temperature, discretized in
such a way that $\beta=\Delta\tau L$. After the application of the
Hubbard-Stratonovich transformation \cite{hubbard59,
stratonovich58} the fermionic degrees of freedom in Eq.
(\ref{zini}) may be traced out, and we arrive at
\begin{eqnarray}
Z_{\textrm{GC}}&=&\sum_{\{\sigma\}}\prod_{\alpha}\det [1+B_{L}(\alpha)B_{L-1}(\alpha)\cdots
B_{1}(\alpha)]\nonumber\\
&=&\sum_{\{\sigma\}}\det O(\{\sigma\},\mu)_{\uparrow}\det O(\{\sigma\},\mu)_{\downarrow}.
\label{zfinal}
\end{eqnarray}
The $B_l$ matrices are defined as
\begin{eqnarray}
B_{l}(\alpha)&=&e^{-\Delta\tau K/2}e^{V^{\alpha}(l)}e^{-\Delta\tau K/2},\label{bdefine}\\
(K)_{ij}&=&\left\{\begin{array}{cc}
-t_{ij} & \mbox{for $i$,$j$ nearest neighbours},\\
0 & \mbox{otherwise},
\end{array}\right.\\
V_{ij}^{\alpha}(l)&=&\delta_{ij}[\lambda\alpha\sigma_{i}(l)+\mu\Delta\tau],
\label{vls}
\end{eqnarray}
where $\sigma_i(l)=\pm 1$ is the auxiliary Ising spin coupled with
the electrons at lattice site $i$ and time $l\Delta\tau$, and
$\alpha=\pm 1$ corresponds to $\uparrow$ or $\downarrow$ in Eq.
(\ref{zfinal}). In Eq. (\ref{bdefine}) we have used a symmetric
decomposition of the partition function, which produces a much
smaller Trotter error \cite{assaadnic, fye86} compared to the
non-symmetric decomposition.\cite{sugar81a, sugar81b,
hirsch83a83b, white89} The Monte Carlo (MC) weight $P$ is then
given by the product of two determinants in Eq. (\ref{zfinal}),
which is always positive for bipartite lattices at half-filling.
\cite{hirsch85} At low temperatures we use QR factorization to
stabilize the matrix multiplications and inversions. \cite{white89,
golub89} A version (projector QMC or PQMC) of the above procedure
can be used to directly project out the ground state properties
from an initial trial wave function; see Ref. \onlinecite{white89}
for details.

\begin{table*}
  \centering
  \begin{tabular*}{\hsize}{@{\extracolsep{\fill}}|ll|dd|dd|dd|dd|}
    \hline\hline
    Part A  &   &  U=2t   &      &  U=3t & & U=4t & & U=4.5t & \\
    $n$ & $S_z$ & E_n(S_z) & \textrm{sign} & E_n(S_z) & \textrm{sign} & E_n(S_z) & \textrm{sign} & E_n(S_z) & \textrm{sign} \\
    \hline
    57 & 1/2 & -74.535(5) & 0.69 & -64.72(2)  & 0.15 & -56.6(6)^{*}   & 0.02   &      &  \\
    57 & 3/2 & -74.574(4) & 0.81 & -64.79(2)  & 0.25 & -57.1(1)^{*}   & 0.04   &      &  \\
    58 & 0   & -74.290(3) & 0.75 & -64.06(2)  & 0.25 & -55.82(8)^{*}  & 0.04   &    &  \\
    58 & 1   & -74.322(4) & 0.82 & -64.098(9)  & 0.32 & -55.95(5)^{*}  & 0.07  &     &  \\
    59 & 1/2 & -74.080(4) & 0.89 & -63.366(8) & 0.51 & -54.74(2)^{*}  & 0.22   &   &  \\
    59 & 3/2 & -73.104(4) & 0.83 & -62.475(7)  & 0.31 & -54.06(4)^{*}  & 0.06  &    &  \\
    60 & 0   & -73.810(3) & 1.00 & -62.633(3) & 1.00 & -53.091(2) & 0.98    & -48.969(3) & 0.94 \\
    60 & 1   & -72.885(4)^{*} & 0.92 & -61.83(1)^{*}  & 0.44 & -52.30(2)^{*}  & 0.29 &  &         \\
    61 & 1/2 & -72.448(2) & 0.98 & -60.704(3) & 0.82 & -50.542(5) & 0.47    & -46.080(5) & 0.32\\
    61 & 3/2 & -71.547(3)^{*} & 0.89 & -59.957(7)^{*} & 0.46 & -50.21(3)^{*}  & 0.13 &  &  \\
    62 & 0   & -71.043(4) & 0.95 & -58.728(6) & 0.63 & -47.92(1) & 0.22     &  -43.15(4) & 0.10 \\
    62 & 1   & -71.072(2) & 0.98 & -58.756(3) & 0.77 & -47.965(6) & 0.35    & -43.175(9) & 0.18 \\
    63 & 1/2 & -69.649(3) & 0.96 & -56.760(5) & 0.60 & -45.34(2) & 0.17     &  &  \\
    63 & 3/2 & -69.688(4) & 1.00 & -56.802(3) & 0.88 & -45.360(8) & 0.40    &  &  \\
    64 & 0   & -68.227(3) & 0.95 & -54.735(8) & 0.54 & -42.67(5)  & 0.12    &  &  \\
    64 & 1   & -68.252(4) & 0.98 & -54.743(5) & 0.71 & -42.69(2)  & 0.20    &  &  \\
    65 & 1/2 & -66.801(3) & 0.98 & -52.719(7) & 0.70 & -39.96(3)  & 0.17    &  &  \\
    65 & 3/2 & -66.587(5) & 0.96 & -52.505(8) & 0.61 & -39.85(2)  & 0.15    &  &  \\
    66 & 0   & -65.337(4) & 1.00 & -50.638(4) & 0.81 & -37.26(2) & 0.21     &   &  \\
    66 & 1   & -65.115(3) & 0.95 & -50.419(9) & 0.58 & -37.07(4)& 0.13      &   &  \\
    \hline
    \hline
    Part B & & U=2t & & U=3t & & U=4t & & U=4.5t & \\
     $n$  & $S_z$ & \Delta_b(n) & & \Delta_b(n) & & \Delta_b(n) & & \Delta_b(n) & \\
    \hline
    58 & (1/2,3/2,1) & -0.010(8) & & 0.04(3) & & 0.1(2)^{*} & &  & \\
    59 & (0,1,1/2) & 0.028(8) & & 0.00(1) & & -0.06(5)^{*} & &  & \\
    60 & (1/2,1/2,0) & 1.092(6) & & 1.20(1) && 1.42(2)^{*} & &  &\\
    61 & (1,0,1/2) & 0.014(5)& & 0.019(6)& & 0.028(9) & & 0.02(1)& \\
    62 & (3/2,1/2,1)& 0.008(5) & & 0.006(6) & & 0.03(1)& & &\\
    63 & (1,1,3/2)& 0.052(7) && 0.105(7) & & 0.07(2) & &  & \\
    64 & (1/2,3/2,1)& 0.015(8) && -0.04(1) & & 0.05(4) & &  & \\
    65 & (0,1,1/2)& 0.013(7) & & 0.06(1)& & -0.03(5) & &  & \\
  \hline\hline
  \end{tabular*}
  \caption{PQMC calculations on a $C_{60}$ molecule. \textbf{Part A}
  of the table shows the total internal energy $E_n(S_z)$ of a
  $C_{60}$ molecule with $n$ electrons and $z$-component of total spin
  $S_z$. The parameters used in the simulations are $t'=1.2t$,
  $\beta=10/t$, $\Delta\tau=0.0625/t$ (for $U=4t$),
  $\Delta\tau=0.05/t$ (for other $U$ values), $N_m=10^7$. $N_m$ data
  were divided into 10 bins for error estimation. For $n=60,61,62$, we
  have collected more data (between $4\times 10^7$ and $8\times 10^7$
  measurements) for a more accurate comparison between PQMC and
  perturbative results. \textbf{Part B} shows electron (hole) binding
  energies $\Delta_{b}(n)$. As before, the $S_z$ column in this case
  lists the $S_z$ values of three states, involved in the calculation of
  the binding energy, in the order of appearance in
  Eq. (\ref{pairbind}). For example, $\Delta_{\textrm{b}}(58)$ with
  $S_z=(1/2,3/2,1)$ denotes $E_{59}(S_z=1/2)+E_{57}(3/2)-2E_{58}(1)$.
  The data points marked with $*$ were calculated using a
  non-symmetric decomposition in Eq. (\ref{bdefine}). Only limited results were obtained for
  $U=4.5t$ because of the long averaging times required.} \label{c60pqmc}
\end{table*}

Unless indicated otherwise, PQMC and AFQMC calculations were performed
at the temperature fixed by $\beta=10/t$, with imaginary time
discretization $\Delta \tau=0.05/t$. The system was first brought to
thermal equilibrium by performing $N_t$ thermalizations sweeps $S2$,
followed by $N_m$ measurements with a single sweep $S1$\cite{measure}
performed between them.

Since we are interested in non-bipartite molecules such as $C_{60}$,
the MC weights in general are not always positive. In the case of
negative weight $P$, we associate a probability value $|P|$ with it,
and include a sign $S=P/|P|$ in the average: $\langle E\rangle=\langle
ES\rangle/\langle S\rangle$. Now the average $\langle\ldots\rangle$ is
with respect to the probability distribution $|P|$.

Finally, we note that in estimating the statistical error for a
composite quantity $X=X_1+X_2+\ldots+X_n$, we use the standard formula
$\delta X=(\delta X_1^2+\delta X_2^2+\ldots+\delta X_n^2)^{1/2}$, where
$\delta X_i$'s are estimated by the Jackknife method.\cite{newman99}

\section{Application}
\subsection{Comparison to other methods}

In this section, we check our QMC programs against ED and SSE results.
\cite{sandvik91, sandvik9799, sandvik02} Table \ref{c12pqmc}
lists energies and binding energies from both ED and PQMC for a truncated tetrahedron, which has 12
lattice sites and 3 nearest neighbors for each site.  For energies
at different dopings, good agreement is obtained between the two
methods. The largest deviation of PQMC from ED
is found for $n=10$ $(S_z=0)$ (about $0.7\%$ deviation), which might
be due to the relatively low value of the average sign and the incomplete
projection of a nearby singlet excitation. Energy
differences between two different dopings (e.g., $\Delta_{1,0}$)
are in good agreement for the two methods. However, the inaccuracies are magnified when
pair-binding energies are extracted from two already-small energy
differences, although some of these energies are still in good
agreement for the two methods within error bounds, e.g.,
$\Delta_{\textrm{b}}(11)$. The difficulty of extracting
$\Delta_{\textrm{b}}(13)$ from PQMC is possibly because the ground
state with 14 electrons lies in the spin singlet sector (as confirmed by ED), and it
is difficult for PQMC to completely project out the nearby spin
triplet state (first excited state). There is no such problem if
the ground state for two-electron doping is a spin triplet, which,
as we will see below, is exactly the case for $C_{60}$. From the
good agreement between ED and PQMC, we conclude that the
discretization error caused by $\Delta\tau=0.05/t$ is sufficiently
small. We also find that the projection factor
$\beta=10/t$ is large enough to project out the ground state from
an initial trial state. We will use these values of $\Delta\tau$
and $\beta$ in our AFQMC and PQMC simulations of the $C_{60}$
molecule.

In Table \ref{c60afqmc}, we check our grand canonical simulation
program (AFQMC), against ED (at $U=0$) and SSE (at $U=4t$) on a $C_{60}$ molecule. Again we see good
agreement in both the density $n$ and energy $E$ calculations
among these methods. For $U=0$, the AFQMC results are exactly the
same as the exact diagonalization results. This is because at
$U=0$, there is no coupling between electrons and the auxiliary
Ising field; the Ising field is wiped out completely and the
electrons cannot feel the existence of Ising spins. The simulation
at $U=0$ also shows that the discretization error is absent in
AFQMC. In both $U=0$ and $U=4t$, we have set the temperature
$T=2t$ to avoid a severe sign problem in the SSE simulation.
Because SSE does not suffer from the discretization error, we again
confirm that $\Delta\tau=0.05/t$ in AFQMC is sufficiently small to avoid any systematic
discrepancy.

The results in Table \ref{c12pqmc} for simulations on a truncated
tetrahedron molecule show that PQMC results are in good agreement
with ED. The systematic discretization error
caused by $\Delta\tau$ is reasonably small; thus the
extrapolation to $\Delta\tau=0$ is unnecessary.

\subsection{Application to $C_{60}$ molecule}
In this section, we discuss the QMC results for a $C_{60}$
molecule. Figure \ref{c60gcs} shows the results of an AFQMC
simulation of the Hubbard model on a $C_{60}$ molecule at various
chemical potentials $\mu$.
\begin{figure}
  \begin{tabular}{c}
    \resizebox{75mm}{!}{\includegraphics{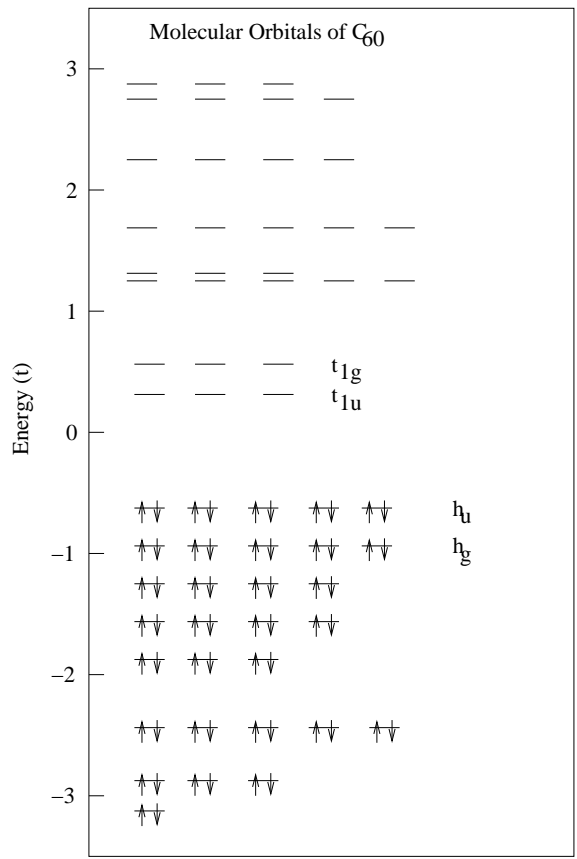}} \\
  \end{tabular}
  \caption{Huckel energy level diagram for the neutral $C_{60}$ molecule. The
lowest 30 levels are doubly occupied. The energy level scale is
drawn according to the exact diagonalization of the non-interacting
on-site Hubbard Hamiltonian, i.e., $U=0$. The energy level labels
are from those of the icosahedral group, which is the symmetry
group of a $C_{60}$ molecule. The LUMO band is labelled by
$t_{1u}$, and HOMO by $h_u$. We will consider doping of LUMO and
HOMO for a discussion of Hund's rule.}\label{c60band}
\end{figure}
At half-filling, unlike the bipartite 2D square lattice, the QMC
simulation has a slight sign problem due to the pentagon
frustration in the $C_{60}$ geometry; see panels (c) and (d) in
Figure \ref{c60gcs}. From (c) and (d) we also see that hole dopings
have a worse sign problem than electron dopings. As expected, the
compressibility \cite{moreo90}
$\kappa\equiv\frac{1}{n^2}\frac{dn}{d\mu}\sim 0$ around $\mu=0$.

Table \ref{c60pqmc} lists the PQMC results for $2\le U/t\le 4.5$ for the
$C_{60}$ molecule. It can be seen that hole doping causes a more
severe sign problem than the electron doping, which is
consistent with the behavior in Figure \ref{c60gcs}. In Part A of
the table we see that reasonably accurate results can be obtained
for $U=2t$, $3t$ and $4t$. The sign problem quickly becomes severe
for $U>4t$, as is evident in the data presented for $U=4.5t$, where the
average sign is only 0.18 for 62 electrons.  For $U=5t$ and 62 electrons,
the average sign is 0.08 and it is not possible to extract a reliable
binding energy.  

\begin{figure}
  \begin{tabular}{c}
    \resizebox{70mm}{!}{\includegraphics{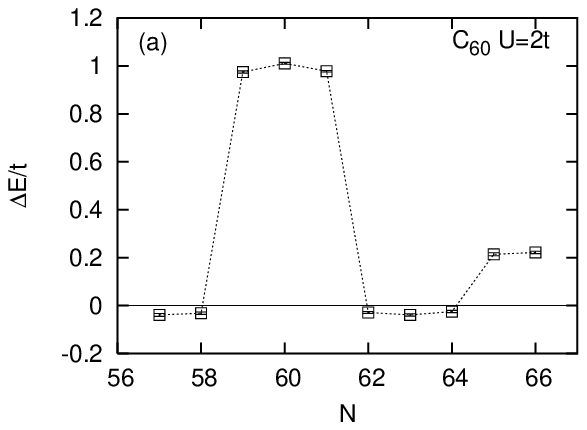}} \\
    \resizebox{70mm}{!}{\includegraphics{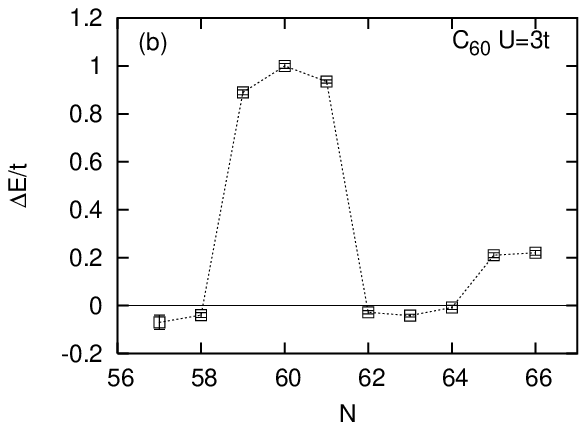}} \\
    \resizebox{70mm}{!}{\includegraphics{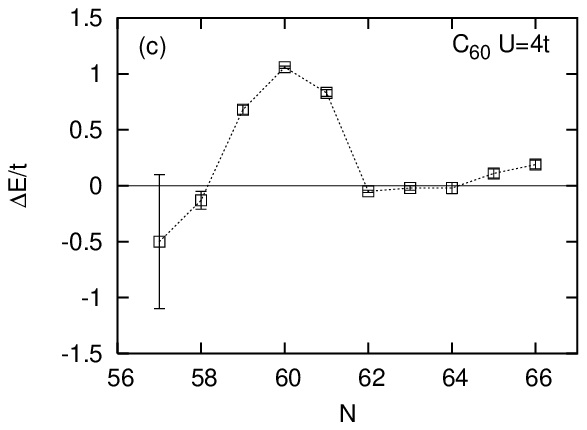}} \\
  \end{tabular}
  \caption{Comparison of the PQMC spin configuration of a $C_{60}$ molecule at various dopings
  with Hund's rule. $\Delta E$ is defined as the energy difference between two total
  spin $z$ component sectors, e.g., $\Delta E(60)=E_{60}(S_z=1)-E_{60}(S_z=0)$ for the
  neutral molecule and $\Delta E(61)=E_{61}(S_z=3/2)-E_{61}(S_z=1/2)$ for one-electron
  doping. A positive $\Delta E$ at fillings $n=59, 60, 61$ can be understood in the
  non-interacting picture in Figure \ref{c60band}, and a negative $\Delta E$ at fillings
  $n=57, 58, 62, 63, 64$ are in agreement with Hund's rule. $n=65, 66$ can again be
  explained with Figure \ref{c60band}. See text for discussions. The dotted lines connecting
  MC points are only guides to the eye.}\label{c60hund}
\end{figure}
From Table \ref{c60pqmc}A we can calculate the energy difference
between different total spin sectors at the same doping to compare
the ground state spin configuration with Hund's rule. Let us first
discuss the non-interacting single electron energy levels of a
$C_{60}$ molecule. At half-filling, 60 electrons move
independently in ``$\pi$'' molecular orbitals formed from the 60
$p_z$ atomic orbitals of the 60 carbon atoms. An exact
diagonalization of the non-interacting Hamiltonian gives 60 energy
levels, of which the lowest 30 levels are occupied at half
filling. The lowest unoccupied molecular orbitals (LUMO) are 3-fold
degenerate. The highest occupied molecular orbitals (HOMO) are 5-fold degenerate. 
The detailed energy levels of the neutral $C_{60}$
molecule are shown in Figure \ref{c60band}. The energy gap between HOMO
and LUMO is $1.04t$ from this exact diagonalization for
non-interacting neutral $C_{60}$.

We calculate the energy difference $\Delta E$ of different total
$S_z$ values at the same filling as shown in Figure \ref{c60hund}.
As a severe sign problem occurs for $U=5t$, the statistical errors
are too large for a definite discussion of Hund's rule. Thus, we
mainly discuss results in Figure \ref{c60hund} (a)-(c). At $n=59,
60$ and 61, $\Delta E$ is positive, which can be explained by the
non-interacting energy levels in Figure \ref{c60band}. For
example, for $n=59$ and total spin $S_z=1/2$, we need to flip one
electron from spin down to spin up in order to get $S_z=3/2$,
which means we need to excite one spin-down electron from the HOMO
band to the LUMO band, with an energy cost of $1.04t$. The same
explanation applies to $n=60$ and 61. At $n=57, 58, 62, 63, 64$,
$\Delta E$ is close to zero which means that the two different
values of $S_z$ are part of the same multiplet. For these cases,
in agreement with Hund's rule, the electrons tend to occupy the
degenerate HOMO or LUMO in a way that maximizes the total spin.

Note that the small differences in energy for fixed $n$ and
different $S_z$ within a multiplet must result from small
admixtures of excited states. For example, if the ground state is
a triplet, as for $n=62$, and if the first excited state is a
singlet, then the $S_z=0$ state may contain a small admixture of
the excited singlet state, while the $S_z=1$ state will not. This
will result in the $S_z=0$ state having a slightly higher energy
than the $S_z=1$ state, providing a measure of the effectiveness
of the projection within this multiplet.

At $n=65, 66$, the positive $\Delta E$ can again be explained by
the single electron picture. A spin-down electron in the $t_{1u}$
band must flip its spin and be excited to the $t_{1g}$ band,
with an energy cost of $0.2603t$ for the non-interacting Hamiltonian. This
is consistent with Figure \ref{c60hund} for $n=65$ and 66.

In Table \ref{c60pqmc}B we calculate the pair binding energy
$\Delta_{\textrm{b}}(n)$ at various dopings. In these calculation, we have used
the lowest energy of the different total spin $z$ states for a given
$n$. Figure \ref{perturb61} shows a comparison between the
perturbation calculations of Refs. \onlinecite{kivelson91a,
kivelson91b} and PQMC calculations. There is no indication from the
PQMC calculations of a bound singlet state for $U>3.3t$ as suggested
by perturbation theory, represented by the solid line in
Figure \ref{perturb61}. Instead, the ground state for 62 electrons is a
triplet over the entire range of parameters studied, in agreement with
Hund's rule.  Furthermore, we find that this triplet state is unbound
for $U/t\le 4.5$.  Unfortunately, the sign problem precludes us from
studying larger values of $U$ using PQMC.  We have also checked the
energy difference obtained from PQMC with the results from imaginary
chemical potential simulations, which shows good agreement and will
be presented elsewhere.\cite{note1}

\section{Conclusions}
We have performed extensive QMC simulations on a single $C_{60}$
molecule. The PQMC simulation calculates internal energies at various
fillings, and shows that Hund's rule is well obeyed. In contradiction
to the perturbation theory result, \cite{kivelson91a, kivelson91b,
ostlund} we find no singlet pair binding (i.e., no negative
$\Delta_{\textrm{b}}(n)$) for the parameter ranges explored ($U=2t,
3t, 3.5t, 4t, 4.5t$, $t'=1.2t$).  Therefore, a purely electronic
attractive interaction, originating from the one band Hubbard model
with on-site Coulomb interaction, seems unlikely. This main result is 
presented in Figure \ref{perturb61}.

\begin{acknowledgments}
We gratefully acknowledge the support of this project by Natural Sciences and Engineering Research
Council (Canada), The Canadian Institute for Advanced Research (CIAR),
CFI and SHARCNET. All the calculations were carried out at SHARCNET supercomputing facilities at
McMaster University.
\end{acknowledgments}


\begin{thebibliography}{99}
\bibitem{note1} Methodology and results for the imaginary chemical
potential method applied to Hubbard molecules will be published
separately.
\bibitem{measure} To describe our computational procedure, we define
an AFQMC or PQMC sweep $S1$ as an attempt to flip the auxiliary
spins on the $N$ spatial sites, belonging to a single time slice,
and $S2$ as an attempt to flip all the spins on all time slices.
\end{thebibliography}
\end{document}